\documentclass{template/Interspeech2024}

\interspeechcameraready

\usepackage{bm}
\usepackage{diffcoeff}
\usepackage{pgfplots} 
\pgfplotsset{compat=1.18} 
\usepackage{tikz}
\usetikzlibrary{positioning}
\usepackage{placeins} 
\usepackage{multirow}
\usepackage{cite}
\usepackage{stfloats}
\usepackage{hyperref}

\makeatletter
\renewcommand{\section}{\@startsection
  {section}
  {1}
  {}
  {-\baselineskip}
  {0.25\baselineskip}
  {}}

\renewcommand{\subsection}{\@startsection
  {subsection}
  {2}
  {}
  {-\baselineskip}
  {0.25\baselineskip}
  {}}

\renewcommand{\subsubsection}{\@startsection
  {subsubsection}
  {3}
  {}
  {-\baselineskip}
  {0.25\baselineskip}
  {}}
\makeatother

\renewenvironment{thebibliography}[1]{
  \begin{oldthebibliography}{#1}
    \setlength{\itemsep}{2.0pt} 
    \setlength{\parskip}{1.0pt}
}
{
  \end{oldthebibliography}
}
\def\ninept{\def\baselinestretch{.898}\let\normalsize\small\normalsize}
\title{Schrödinger Bridge for Generative Speech Enhancement}
\name{Ante}{Jukić}
\name{Roman}{Korostik}
\name{Jagadeesh}{Balam}
\name{Boris}{Ginsburg}
\address{NVIDIA, USA}
\email{\{ajukic, rkorostik, jbalam, bginsburg\}@nvidia.com}
\begin{document}
\ninept
%
\maketitle
\begin{abstract}
This paper proposes a generative speech enhancement model based on Schr\"{o}dinger bridge (SB).
The proposed model is employing a tractable SB to formulate a data-to-data process between the clean speech distribution and the observed noisy speech distribution.
The model is trained with a data prediction loss, aiming to recover the complex-valued clean speech coefficients, and an auxiliary time-domain loss is used to improve training of the model.
The effectiveness of the proposed SB-based model is evaluated in two different speech enhancement tasks: speech denoising and speech dereverberation.
The experimental results demonstrate that the proposed SB-based outperforms diffusion-based models in terms of speech quality metrics and ASR performance, e.g., resulting in relative word error rate reduction of 20\% for denoising and 6\% for dereverberation compared to the best baseline model.
The proposed model also demonstrates improved efficiency, achieving better quality than the baselines for the same number of sampling steps and with a reduced computational cost.
\end{abstract}

\section{Introduction}
Recordings of speech signals are frequently corrupted by environmental noise, undesired sounds and room reverberation.
The undesired signal components may impair the quality or intelligibility for human or machine listeners~\cite{beutelmann2006prediction,yoshioka2012making}.
The goal of speech enhancement (SE) in such scenarios is to recover the clean speech signal from a corrupted recording.

Typically, SE methods exploit statistical properties of the desired speech signal and the undesired disturbance signal.
Classical model-based SE methods rely on a priori knowledge of the statistical properties of either one or both signals~\cite{hendriks2013dft, gerkmann2018spectral}.
Data-driven SE methods typically use machine learning (ML) models to learn signal properties from training data~\cite{wang2018supervised}.
ML-based SE can be broadly divided into predictive and generative models.
Predictive models aim to provide an estimate of the clean signal from the noisy signal, e.g., using an ML model to estimate real- or complex-valued spectral masks~\cite{wang2014training,williamson2015complex}, coefficients~\cite{xu2014regression, han2015learning} or the time domain signal~\cite{luo2018tasnet}.
Generative models aim to model the distribution of the clean signal given the noisy signal, e.g., using variational autoencoders~\cite{kingma2013auto,leglaive2018variance} or generative adversarial networks~\cite{pascual2017segan}.

Recently, several diffusion-based generative models for SE have been proposed~\cite{lu2021study, lu2022conditional, welker22speech, richter_2023_sgmse, lemercier2023storm, lemercier2023analysing, scheibler2023diffusion, guo2023variance, kamo23_interspeech, lay_2023_reducing}.
In general, diffusion-based models are based on two processes between the clean speech distribution and the noisy signal prior distribution~\cite{yang2023diffusion}.
The forward process transforms the clean data into a known prior distribution, and the reverse process starts from the prior distribution and generates an estimate of the clean speech.
A neural network model is trained to guide the reverse process.
In~\cite{lu2021study}, spectrogram of the noisy signal was used as a conditioner for the neural model, but only additive Gaussian noise was considered.
A conditional diffusion model was proposed in~\cite{lu2022conditional} to handle non-Gaussian noise, with the diffusion process conditioned on the noisy signal.
An alternative diffusion process in the short-time Fourier transform (STFT) domain was proposed in~\cite{richter_2023_sgmse}, enabling generative training of the model without any assumptions on the noise distribution.
The proposed model operates on complex-valued time-frequency (TF) coefficients, enabling the neural model to learn the TF structure of the signal, and it has shown to be effective for speech denoising and dereverberation~\cite{lemercier2023analysing}.
However, the model may produce hallucinations in adverse scenarios, resulting in vocalization or breathing artifacts in extreme noise or during spech absence~\cite{richter_2023_sgmse,lemercier2023storm}.
Prior mismatch of the diffusion process was reduced using modified objectives in~\cite{scheibler2023diffusion,kamo23_interspeech} and a modified forward process was used in~\cite{lay_2023_reducing}.
A hybrid model, combining a predictive model and a diffusion-based generative model, was proposed in~\cite{lemercier2023storm} to improve the robustness and reduce the computational complexity of the sampling process.

This paper proposes a generative SE model based on Schr\"{o}dinger bridge (SB)~\cite{schrodinger1932theorie, debortoli2021diffusion, chen2021likelihood, bunne2023schrodinger, chen_2023_sb}.
As opposed to diffusion models, which describe data-to-noise process, the SB describes a data-to-data process.
We consider a special case of SB where the clean speech and the noisy signal are considered as paired data~\cite{chen_2023_sb}.
The contribution of this work is threefold.
Firstly, we propose a SE model based on Schr\"{o}dinger bridge, using a SB for paired data~\cite{chen_2023_sb}.
As opposed to the noisy prior in diffusion-based models, the SB model results in a reverse process starting exactly from the observed noisy data.
Secondly, we propose to combine the data prediction loss with an auxiliary loss to improve the performance of the proposed SB model.
Thirdly, we demonstrate the effectiveness of the proposed approach in two different SE tasks: speech denoising and speech dereverberation.
The results in terms of speech quality metrics and ASR performance demonstrate the effectiveness of the proposed approach, outperforming diffusion-based SE models at a lower computational complexity.
\section{Background}
\label{sec:background}
Assuming a single static speech source, the signal captured by a single microphone can be modeled as $\underline{\mathbf{y}} = \underline{\mathbf{h}} \ast \underline{\mathbf{x}} + \underline{\mathbf{n}}$, where $\underline{\mathbf{h}}$ is the time-domain impulse response, $\underline{\mathbf{x}}$ is the clean speech signal, and $\underline{\mathbf{n}}$ is the additive noise signal.
The goal of SE is to estimate the clean speech signal $\hat{\underline{\mathbf{x}}} \in \mathbb{R}^N$ from the microphone signal $\underline{\mathbf{y}} \in \mathbb{R}^N$.
In the following, $\mathbf{x} = \mathcal{A} \left( \underline{\mathbf{x}} \right) \in \mathbb{C}^{D}$ denotes a vector of complex-valued coefficients obtained from the time-domain signal $\underline{\mathbf{x}}$.
The analysis transform $\mathcal{A}$ is a composition of the STFT transform $\mathcal{F}$ followed by scaling and compression, i.e.,
$\mathcal{A} \left( \mathbf{x} \right) = b |\mathcal{F} \left( \underline{\mathbf{x}} \right) |^a \mathrm{e}^{j \angle{\mathcal{F} \left( \underline{\mathbf{x}} \right) }}$,
with element-wise operations, magnitude $|.|$ and angle $\angle\left(.\right)$, compression coefficient $a \in \left( 0, 1 \right]$, and scale coefficient $b > 0$~\cite{richter_2023_sgmse}.

\subsection{Score-based diffusion for speech enhancement}
\label{subsec:score_based_diffusion}
Score-based diffusion models~\cite{song2019generative, song_2021_score} are based on a continuous-time diffusion process defined by a forward stochastic differential equation (SDE)
\begin{equation}
  \label{eq:forward_sde}
  \dl{\mathbf{x}_t} = \mathbf{f} \left( \mathbf{x}_t, t\right) \dl{t} + g(t) \dl{\mathbf{w}_t} , \quad \mathbf{x}_0 = \mathbf{x}  ,
\end{equation}
where $t \in \left[ 0, T \right]$ is the current time for the process, $\mathbf{x}_t \in \mathbb{C}^D$ is the state of the process, $\mathbf{f}$ is a vector-valued drift, $g$ is a scalar-valued diffusion coefficient, and $\mathbf{w}_t$ is the standard Wiener process.
The corresponding reverse SDE can be expressed as~\cite{song_2021_score}
\begin{equation}
  \label{eq:reverse_sde}
  \dl{\mathbf{x}_t} = \left[ \mathbf{f} \left( \mathbf{x}_t, t\right) - g^2(t) \nabla \log p_t(\mathbf{x}_t)  \right] \dl{t} + g(t) \dl{\bar{\mathbf{w}}_t} ,
\end{equation}
where $\nabla \log p_t(\mathbf{x}_t)$ is the score function of the marginal distribution $p_t$ at time $t$, and $\bar{\mathbf{w}}_t$ is the reverse-time Wiener process.
In~\cite{welker22speech, richter_2023_sgmse}, the conditional relationship between the clean speech $\mathbf{x}$ and the observed noisy speech $\mathbf{y}$ was directly integrated in the SDE in~\eqref{eq:forward_sde} by using an affine drift term $\mathbf{f} \left( \mathbf{x}_t, t \right) = \gamma \left( \mathbf{y} - \mathbf{x}_t \right)$ with stiffness parameter $\gamma$.
Furthermore, a variance-exploding (VE) diffusion coefficient $g(t) = \sqrt{c} k^{t}$ was used with scale $c > 0$ and base $k > 0$, resulting in an \mbox{Ornstein-Uhlenbeck} SDE with VE (OUVE)~\cite{richter_2023_sgmse}.
The conditional transition distribution $p_{t|0}$ for the state $\mathbf{x}_t$ conditioned on the clean speech $\mathbf{x}$ and the noisy observation $\mathbf{y}$ can be expressed as
\begin{equation}
  \label{eq:diffusion_transition_distribution}
  p_{t|0} = \mathcal{N}_\mathbb{C} \left( \bm{\mu}_x(t), \sigma_x^2(t) \mathbf{I} \right) ,
\end{equation}
where $\mathcal{N}_\mathbb{C}$ is a circularly-symmetric complex Gaussian distribution, and $\mathbf{I}$ is the identity matrix.
The mean $\bm{\mu}_x(t)$ and the variance $\sigma_x^2(t)$ in~\eqref{eq:diffusion_transition_distribution} are defined as
\begin{equation}
  \label{eq:diffusion_mean_variance}
  \bm{\mu}_x \left( t \right) = w_x(t) \mathbf{x} + w_y(t) \mathbf{y} ,
  \enspace\thickspace
  \sigma_x^2 \left( t \right) = \frac{c \left( k^{2t} - \mathrm{e}^{-2 \gamma t} \right)}{2 \left( \gamma + \log k \right)} ,
\end{equation}
with $w_x(t) = \mathrm{e}^{-\gamma t}$, $w_y(t) = 1 - \mathrm{e}^{-\gamma t}$~\cite{richter_2023_sgmse}.
To enable inference using the reverse SDE in~\eqref{eq:reverse_sde}, the score function $\nabla \log p_t(\mathbf{x}_t)$ is estimated using a neural network $s_{\bm{\theta}}$ with parameters $\bm{\theta}$~\cite{song_2021_score, richter_2023_sgmse}.
The score function of the conditional transition distribution $p_{t|0}$ can be computed analytically~\cite{song2019generative, song_2021_score}, resulting in a denoising score matching training objective~\cite{richter_2023_sgmse}
\begin{equation}
  \label{eq:diffusion_score_matching_loss}
  \min_\theta \mathcal{E}_{\left( \mathbf{x}, \mathbf{y} \right), t, \mathbf{z}} \big\Vert \sigma_x(t) s_{\bm{\theta}} \left( \mathbf{x}_t, \mathbf{y}, t \right) - \mathbf{z} \big\Vert_2^2 ,
\end{equation}
where $\mathcal{E}$ is the mathematical expectation, $\mathbf{x}_t = \bm{\mu}_x(t) + \sigma_x(t) \mathbf{z}$ and $\mathbf{z} \sim \mathcal{N}_\mathbb{C} \left( 0, \mathbf{I} \right)$, with the mean and variance in~\eqref{eq:diffusion_mean_variance}.
The parameters $\bm{\theta}$ of the neural network are optimized by minimizing~\eqref{eq:diffusion_score_matching_loss}, with the expectation approximated by sampling $\left( \mathbf{x}, \mathbf{y} \right)$ from the training dataset, $t$ uniformly from $\left[ t_\text{min}, T \right]$ with a small $t_\text{min}$ to avoid numerical issues, and $\mathbf{z}$ from a standard Gaussian distribution~\cite{song2019generative, song_2021_score, richter_2023_sgmse}.
Inference is performed using the reverse SDE in~\eqref{eq:reverse_sde} by using the estimated score $s_\mathbf{\theta} \left( \mathbf{x}_t, \mathbf{y}, t \right)$ and starting from an initial sample $\mathbf{x}_T \sim \mathcal{N}_\mathbb{C} \left( \mathbf{y}, \sigma_x^2(T) \mathbf{I} \right)$.
\section{Proposed model}
\label{sec:proposed_model}
\subsection{Schr\"{o}dinger bridge}
\label{subsec:sb}
We consider a Schr\"{o}dinger bridge~\cite{schrodinger1932theorie, debortoli2021diffusion, chen2021likelihood, bunne2023schrodinger, chen_2023_sb} defined as minimization of the Kullback-Leibler divergence $D_\text{KL}$ between a path measure $p$ and a reference path measure $p_\text{ref}$, subject to the boundary conditions
\begin{equation}
  \label{eq:sb}
  \min_{p \in \mathcal{P}_{\left[0, T\right]}} D_\text{KL} \left(p, p_\mathrm{ref} \right)
  \quad
  \text{s. t.}
  \quad
  p_0 = p_x ,
  \thickspace
  p_T = p_y ,
\end{equation}
where $\mathcal{P}_{\left[0, T\right]}$ is the space of path measures on $\left[0, T\right]$.
Assuming $p_\mathrm{ref}$ is defined by the reference forward SDE in~\eqref{eq:forward_sde}, the SB is equivalent to a pair of forward-backwards SDEs~\cite{chen2021likelihood,chen_2023_sb}
\begin{align}
  \label{eq:sb_forward_sde}
  \dl{ \mathbf{x}_t } &= \left[ \mathbf{f} + g^2(t) \nabla \log \Psi_t \right] \dl{t} + g(t) \dl{\mathbf{w}_t}, \, \mathbf{x}_0 \sim p_x , \\
  \label{eq:sb_reverse_sde}
  \dl{ \mathbf{x}_t } &= \left[ \mathbf{f} - g^2(t) \nabla \log \bar{\Psi}_t \right] \dl{t} + g(t) \dl{\bar{\mathbf{w}}_t}, \, \mathbf{x}_T \sim p_y ,
\end{align}
where scores of $\Psi, \bar{\Psi}_t$ are the optimal forward and reverse drifts, and some function arguments are omitted for brevity.
The marginal distribution of the SB state $\mathbf{x}_t$ can be expressed as $p_t = \bar{\Psi}_t \Psi_t$~\cite{chen2021likelihood, chen_2023_sb}.
While solving the SB is in general intractable, closed-form solutions exist for special cases, e.g., for Gaussian boundary conditions~\cite{bunne2023schrodinger, chen_2023_sb}.
\subsection{Schr\"{o}dinger bridge between paired data}
\label{subsec:sb_paired_data}
Assume a linear drift $\mathbf{f} \left( \mathbf{x}_t \right) = f(t) \mathbf{x}_t$ and Gaussian boundary conditions $p_0 = \mathcal{N}_\mathbb{C} \left( \mathbf{x}, \epsilon_0^2 \mathbf{I} \right)$ and $p_T = \mathcal{N}_\mathbb{C} \left( \mathbf{y}, \epsilon_T^2 \mathbf{I} \right)$ with $\epsilon_T = \mathrm{e}^{\int_0^T f(\tau) \dl{\tau}} \epsilon_0$.
When $\epsilon_0 \to 0$, the SB solution between clean data $\mathbf{x}$ and noisy data $\mathbf{y}$ can be expressed as~\cite{chen_2023_sb}
\begin{equation}
  \label{eq:sb_solution}
  \bar{\Psi}_t = \mathcal{N}_\mathbb{C} \left( \alpha_t \mathbf{x}, \alpha_t^2 \sigma_t^2 \mathbf{I} \right) ,
  \quad
  \Psi_t = \mathcal{N}_\mathbb{C} \left( \bar{\alpha}_t \mathbf{y}, \alpha_t^2 \bar{\sigma}_t^2 \mathbf{I} \right) ,
\end{equation}
with parameters $\alpha_t = \mathrm{e}^{\int_0^t f(\tau) \dl{\tau}}$, $\sigma_t^2 = \int_0^t \frac{g^2(\tau)}{\alpha_\tau^2} \dl{\tau}$, $\bar{\alpha}_t = \alpha_t \alpha_T^{-1}$ and $\bar{\sigma}_t^2 = \sigma_T^2 - \sigma_t^2$.
Therefore, the marginal distribution $p_t = \bar{\Psi}_t \Psi_t$ is also a Gaussian distribution which can be expressed as
\begin{equation}
  \label{eq:sb_marginal_distribution}
  p_t = \mathcal{N}_\mathbb{C} \left( \bm{\mu}_x(t), \sigma_x^2(t) \mathbf{I} \right) .
\end{equation}
The mean $\bm{\mu}_x(t)$ and the variance $\sigma_x^2(t)$ in~\eqref{eq:sb_marginal_distribution} are defined as
\begin{equation}
  \label{eq:sb_mean_variance}
  \bm{\mu}_x(t) = w_x(t) \mathbf{x} + w_y(t) \mathbf{y} ,
  \quad
  \sigma_x^2(t) = \frac{ \alpha_t^2 \bar{\sigma}_t^2 \sigma_t^2 }{ \sigma_T^2 } ,
\end{equation}
with $w_x(t) = \alpha_t \bar{\sigma}_t^2 / \sigma_T^2$, and $w_y(t) = \bar{\alpha}_t \sigma_t^2 / \sigma_T^2$~\cite{chen_2023_sb}.
\begin{table}
  \caption{Noise schedules used for the proposed SB-based SE.}
  \label{table:noise_schedule}
  \vspace{-1.7em}
  \begin{center}
    \begin{tabular}{ccccc}
      \toprule
      Noise schedule & Scaled VP                                                                               & VE \\ \midrule
      $f(t)$         & $-\frac{\beta_0 + \left( \beta_1 - \beta_0 \right) t}{2}$                               & 0 \\ \midrule
      $g^2(t)$       & $c \left[ \beta_0 + (\beta_1 - \beta_0) t \right]$                                      & $c k^{2t}$ \\ \midrule
      $\alpha_t$     & $\mathrm{e}^{-\frac{1}{2} \left( \beta_0 t + \frac{\beta_1 - \beta_0}{2} t^2  \right)}$ & 1 \\ \midrule
      $\sigma_t^2$   & $c \left( \mathrm{e}^{\beta_0 t + \frac{\beta_1 - \beta_0}{2} t^2} - 1 \right)$         & $\frac{c \left( k^{2t} - 1 \right)}{2 \log \left( k \right)}$ \\
      \bottomrule
    \end{tabular}
  \end{center}  
  \end{table}
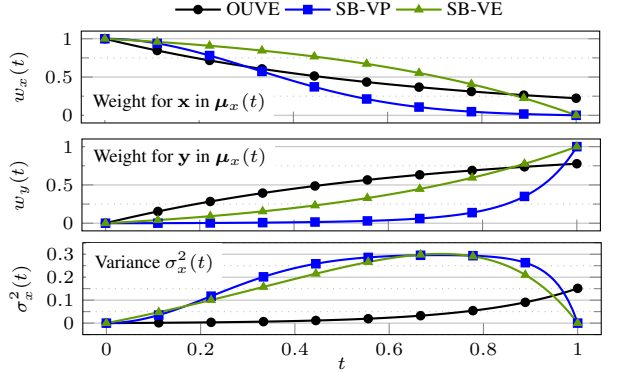
\begin{figure}
  \pgfplotstableread[col sep=comma]{tables/schedule_w_x.csv}\weightx
  \begin{tikzpicture}
    \begin{axis}[
        width=8.4cm,
        height=2.8cm,
        title=Weight for $\mathbf{x}$ in $\bm{\mu}_x(t)$,
        title style={xshift=-6.7em, yshift=-4.4em, fill=white, fill opacity=0.5, text opacity=1.0},
        ylabel={$w_x(t)$},
        ymajorgrids=true,
        yminorgrids=true,
        minor y tick num=1,
        minor grid style={dotted},
        xtick pos=bottom,
        xticklabel=\empty,
        legend pos=north east,
        legend style={legend columns=3, nodes={scale=0.8}, at={(0.5,1.43)}, anchor=north, draw=none},
        title style={font=\scriptsize},
        label style={font=\scriptsize},
        tick label style={font=\scriptsize},
        xtick align=inside,
        xmin=0,
        xmax=1,
        ymin=0,
        ymax=1,
        enlarge x limits = {abs=0.05},
        enlarge y limits = {abs=0.1}
      ]
      \addplot[draw=black, thick, mark=*, mark repeat=11, mark options={scale=0.7, fill=black}] table[x=t, y=ouve]{\weightx};
      \addplot[draw=blue, thick, mark=square*, mark repeat=11, mark options={scale=0.7, fill=blue}] table[x=t, y=sb_vp]{\weightx};
      \addplot[draw={rgb:red,118;green,185;blue,0}, thick, mark=triangle*, mark repeat=11, mark options={scale=0.7, fill={rgb:red,118;green,185;blue,0}}] table[x=t, y=sb_ve]{\weightx};
      \legend{OUVE, SB-VP, SB-VE}
    \end{axis}
  \end{tikzpicture}

  \vspace{-0.4em}

  \pgfplotstableread[col sep=comma]{tables/schedule_w_y.csv}\weighty
  \begin{tikzpicture}
    \begin{axis}[
        width=8.4cm,
        height=2.8cm,
        title=Weight for $\mathbf{y}$ in $\bm{\mu}_x(t)$,
        title style={xshift=-6.7em, yshift=-2.2em, fill=white, fill opacity=0.5, text opacity=1.0},
        ylabel={$w_y(t)$},
        ymajorgrids=true,
        yminorgrids=true,
        minor y tick num=1,
        minor grid style={dotted},
        xtick pos=bottom,
        xticklabel=\empty,
        title style={font=\scriptsize},
        label style={font=\scriptsize},
        tick label style={font=\scriptsize},
        xtick align=inside,
        xmin=0,
        xmax=1,
        ymin=0,
        ymax=1,
        enlarge x limits = {abs=0.05},
        enlarge y limits = {abs=0.1}
      ]
      \addplot[draw=black, thick, mark=*, mark repeat=11, mark options={scale=0.7, fill=black}] table[x=t, y=ouve]{\weighty};
      \addplot[draw=blue, thick, mark=square*, mark repeat=11, mark options={scale=0.7, fill=blue}] table[x=t, y=sb_vp]{\weighty};
      \addplot[draw={rgb:red,118;green,185;blue,0}, thick, mark=triangle*, mark repeat=11, mark options={scale=0.7, fill={rgb:red,118;green,185;blue,0}}] table[x=t, y=sb_ve]{\weighty};
    \end{axis}
  \end{tikzpicture}

  \vspace{-0.4em}

  \pgfplotstableread[col sep=comma]{tables/schedule_sigma_x_sq.csv}\sigmaxsq
  \begin{tikzpicture}
    \begin{axis}[
        width=8.4cm,
        height=2.8cm,
        title=Variance $\sigma_x^2(t)$,
        title style={xshift=-7.8em, yshift=-2.3em, fill=white, fill opacity=0.5, text opacity=1.0},
        xlabel={$t$},
        xlabel style={yshift=1.3ex},
        ylabel={$\sigma_x^2(t)$},
        ylabel style={yshift=-0.18ex},
        ymajorgrids=true,
        yminorgrids=true,
        minor y tick num=1,
        minor grid style={dotted},
        xtick pos=bottom,
        title style={font=\scriptsize},
        label style={font=\scriptsize},
        tick label style={font=\scriptsize},
        xtick align=inside,
        xmin=0,
        xmax=1,
        ymin=0,
        ymax=0.3,
        enlarge x limits = {abs=0.05},
        enlarge y limits = {abs=0.05}
      ]
      \addplot[draw=black, thick, mark=*, mark repeat=11, mark options={scale=0.7, fill=black}] table[x=t, y=ouve]{\sigmaxsq};
      \addplot[draw=blue, thick, mark=square*, mark repeat=11, mark options={scale=0.7, fill=blue}] table[x=t, y=sb_vp]{\sigmaxsq};
      \addplot[draw={rgb:red,118;green,185;blue,0}, thick, mark=triangle*, mark repeat=11, mark options={scale=0.7, fill={rgb:red,118;green,185;blue,0}}] table[x=t, y=sb_ve]{\sigmaxsq};
    \end{axis}
  \end{tikzpicture}
\vspace{-0.9em}
\caption{State $\mathbf{x}_t$ mean $\bm{\mu}_x(t)$ and variance $\sigma_x^2(t)$ for OUVE noise schedule in~\eqref{eq:diffusion_mean_variance} and SB noise schedules in~\eqref{eq:sb_mean_variance}, $t \in \left[0, 1\right]$.}
\label{fig:noise_schedule}
\end{figure}

Several different noise schedules defined by $f(t)$ and $g(t)$ have been considered in~\cite{chen_2023_sb}.
We use a variance preserving (VP) schedule with an additional scaling parameter $c$ for the diffusion coefficient to match the variance of the diffusion-based models.
Similarly as in Section~\ref{subsec:score_based_diffusion}, we also consider a VE diffusion coefficient with the drift term set to zero.
Table~\ref{table:noise_schedule} includes parametrization of the VP and VE noise schedules used here and expressions for their parameters $\alpha_t$ and $\sigma_t^2$.
Figure~\ref{fig:noise_schedule} shows the noise schedules used in Section~\ref{sec:experiments} in terms of mean and variance evolution over time.
As noted in~\cite{lay_2023_reducing}, the \mbox{OUVE} schedule exhibits the mismatch of the mean at the final time $t=1$.
Due to the constraints in~\eqref{eq:sb}, the mean for both SB schedules exactly interpolates between the clean data $\mathbf{x}$ at $t=0$ and the noisy data $\mathbf{y}$ at $t=1$.
Note that the VE and VP schedules have a similar variance evolution, but different mean evolution.
\subsection{Model training}
\label{subsec:model_training}
As noted in~\cite{chen_2023_sb}, the backbone neural model can be trained to match the score $\nabla \log p_t$, to predict the noise using $\nabla \log \bar{\Psi}_t$, or to predict the data $\mathbf{x}$.
We consider the data prediction loss, since it performed well in~\cite{chen_2023_sb}.
Additionally, we propose to use an auxiliary loss $\mathcal{L}_\text{aux}$ to improve the estimate of the model, resulting in the following training objective
\begin{equation}
  \label{eq:data_prediction_loss}
  \min_\theta \mathcal{E}_{\left( \mathbf{x}, \mathbf{y} \right), t, \mathbf{z}} \frac{1}{D} \lVert \hat{\mathbf{x}}_\theta(t) - \mathbf{x} \rVert_2^2 + \lambda \mathcal{L}_\text{aux} \left( \underline{\hat{\mathbf{x}}}_\theta(t), \underline{\mathbf{x}} \right) ,
\end{equation}
where $\hat{\mathbf{x}}_\theta(t) = d_\theta(\mathbf{x}_t, \mathbf{y}, t)$ is the current estimate using a neural network $d_\theta$, $\hat{\underline{\mathbf{x}}}_\theta(t) = \mathcal{A}^{-1} \left( \hat{\mathbf{x}}_\theta(t) \right)$ is the corresponding time-domain signal, and $\lambda > 0$ is a tradeoff parameter.
Using $\lambda=0$ recovers the original data prediction loss.
In Section~\ref{subsec:results}, we report results using the $\ell_1$-norm $\mathcal{L}_\text{aux} \left( \hat{\underline{\mathbf{x}}}, \underline{\mathbf{x}} \right) = \frac{1}{N} \lVert \hat{\underline{\mathbf{x}}} - \underline{\mathbf{x}} \rVert_1$.
Note that we obtained a similar performance using the negative soft-thresholded \mbox{SI-SDR}~\cite{wisdom2020unsupervised}.
\subsection{Inference}
\label{subsec:inference}
\begin{table*}
\caption{Samplers for SB-based SE: solution $\mathbf{x}_t$ at time $t \in \left[ 0, \tau \right]$ given an initial value $\mathbf{x}_\tau$ and $\mathbf{z} \sim \mathcal{N}_\mathbb{C} \left( \mathbf{0}, \mathbf{I} \right)$.}
\vspace{-1.9em}
\label{table:sb_samplers}
\begin{center}
  \setlength\tabcolsep{10.0pt} 
  \resizebox{\textwidth}{!}{
    \begin{tabular}{cc}
      \toprule
      SDE sampler (SB-SDE) & ODE sampler (SB-ODE) \\ \midrule
      $\mathbf{x}_t = \frac{\alpha_t \sigma_t^2}{\alpha_\tau \sigma_\tau^2} \mathbf{x}_\tau + \alpha_t \left( 1 - \frac{\sigma_t^2}{\sigma_\tau^2} \right) \hat{\mathbf{x}}_\theta(\tau) + \alpha_t \sigma_t \sqrt{ 1 - \frac{\sigma_t^2}{\sigma_\tau^2} } \mathbf{z}$ & $\mathbf{x}_t = \frac{\alpha_t \sigma_t \bar{\sigma}_t}{\alpha_\tau \sigma_\tau \bar{\sigma}_\tau} \mathbf{x}_\tau + \frac{\alpha_t}{\sigma_T^2} \left( \bar{\sigma}_t^2 - \frac{\bar{\sigma}_\tau \sigma_t \bar{\sigma}_t}{\sigma_\tau} \right) \hat{\mathbf{x}}_\theta(\tau) + \frac{\alpha_t}{\alpha_T \sigma_T^2} \left( \sigma_t^2 - \frac{\sigma_\tau \sigma_t \bar{\sigma}_t}{\bar{\sigma}_\tau} \right) \mathbf{y}$ \\
      \bottomrule
    \end{tabular}
  }
\end{center}
\end{table*}

\FloatBarrier
The reverse SDE in~\eqref{eq:sb_reverse_sde} can be expressed in terms of the current state $\mathbf{x}_t$ and the current neural estimate $\hat{\mathbf{x}}_\theta(t)$ by computing $\nabla \log \bar{\Psi}_t$ from~\eqref{eq:sb_solution} and replacing $\mathbf{x}$ with $\hat{\mathbf{x}}_\theta (t)$~\cite{chen_2023_sb}.
Given an initial value $\mathbf{x}_\tau$ at time $\tau>0$, the solution of the resulting bridge SDE~\eqref{eq:sb_reverse_sde} at time $t \in \left[ 0, \tau \right]$ can be obtained using first-order discretization in Table~\ref{table:sb_samplers}.
Similarly, a probability flow ordinary differential equation (ODE) formulation can be used to solve the bridge ODE~\cite{chen_2023_sb}, resulting in an ODE sampler in Table~\ref{table:sb_samplers}.
For both the samplers in Table~\ref{table:sb_samplers}, the reverse process starts from $\mathbf{x}_T = \mathbf{y}$, and the final estimate $\hat{\mathbf{x}} = \mathbf{x}_0$ is obtained after a number of steps.
The time-domain output signal is obtained by inverting the analysis transform as $\underline{\hat{\mathbf{x}}} = \mathcal{A}^{-1} \left( \hat{\mathbf{x}} \right)$.
\section{Experiments}
\label{sec:experiments}
%
\begin{table}[t]
\caption{Speech denoising performance on \mbox{WSJ0-CHiME3}. Values are reported as mean ± standard deviation.}
\vspace{-2.0em}
\label{table:enh}
\begin{center}
  \setlength\tabcolsep{3.0pt} 
  \resizebox{\columnwidth}{!}{
    \begin{tabular}{cccccc}
      \toprule
      Signal      & PESQ~$\uparrow$               & ESTOI~$\uparrow$              & SI-SDR/dB~$\uparrow$          & WER/\%~$\downarrow$ \\ \midrule
      Clean       & --                            & --                            & --                            &          3.03 \\
      Unprocessed &          1.35 ± 0.30          &          0.63 ± 0.18          &           4.0 ± 5.8           &         12.18 \\ \midrule
      NCSN++      &          2.18 ± 0.65          & \textbf{0.88} ± \textbf{0.09} & \textbf{16.1} ± \textbf{4.5}  &          5.39 \\
      SGMSE+      &          2.28 ± 0.60          &          0.85 ± 0.11          &          13.1 ± 4.9           &          9.52 \\
      StoRM       &          2.53 ± 0.60          &          0.87 ± 0.09          &          14.8 ± 4.3           &          5.39 \\ \midrule
      SB-VP       & \textbf{2.62} ± \textbf{0.56} & \textbf{0.88} ± \textbf{0.07} &          14.9 ± 4.3           &  \textbf{4.69} \\
      SB-VE       &          2.58 ± 0.53          & \textbf{0.88} ± \textbf{0.07} &          14.7 ± 4.2           &          5.10 \\
      \bottomrule
    \end{tabular}
  }
\end{center}
\end{table}

%
\begin{table}[t]
\caption{Speech dereverberation performance on \mbox{WSJ0-Reverb}. Values are reported as mean ± standard deviation.}
\vspace{-2.0em}
\label{table:der}
\begin{center}
  \setlength\tabcolsep{3.0pt} 
  \resizebox{\columnwidth}{!}{
    \begin{tabular}{cccccc}
      \toprule
      Signal      & PESQ~$\uparrow$               & ESTOI~$\uparrow$              & SI-SDR/dB~$\uparrow$        & WER/\%~$\downarrow$ \\ \midrule
      Clean       & --                            & --                            & --                          &         3.64 \\
      Unprocessed &          1.29 ± 0.13          &          0.44 ± 0.11          &         -9.5 ± 6.3          &         8.29 \\ \midrule
      NCSN++      &          2.00 ± 0.45          &          0.83 ± 0.06          &          5.2 ± 4.2          &         6.45 \\
      SGMSE+      &          2.34 ± 0.43          &          0.82 ± 0.07          &          0.0 ± 8.9          &         5.84 \\
      StoRM       &          2.52 ± 0.41          &          0.85 ± 0.05          &          5.6 ± 4.3          & \textbf{4.69} \\ \midrule
      SB-VP       &          2.26 ± 0.46          &          0.80 ± 0.08          &          4.1 ± 3.9          &         8.62 \\
      SB-VE       & \textbf{2.68} ± \textbf{0.41} & \textbf{0.87} ± \textbf{0.05} & \textbf{6.6} ± \textbf{3.7} &         5.91 \\
      \bottomrule
    \end{tabular}
  }
\end{center}
\end{table}

%
\begin{table*}
\caption{Performance of \mbox{SB-VE} on \mbox{WSJ0-CHiME3} and \mbox{WSJ0-Reverb} trained with $\mathcal{L}_\text{aux}$ and using different samplers for inference.}
\vspace{-2.1em}
\label{table:aux_loss}
\begin{center}
  \setlength\tabcolsep{3.0pt} 
  \resizebox{\textwidth}{!}{
    \begin{tabular}{cccccccccc}
      \toprule
                                                        &         & \multicolumn{4}{c}{\mbox{WSJ0-CHiME3}} & \multicolumn{4}{c}{\mbox{WSJ0-Reverb}} \\ \cmidrule(lr){3-6}\cmidrule(lr){7-10}
      $\mathcal{L}_\text{aux}$                          & Sampler & PESQ~$\uparrow$               & ESTOI~$\uparrow$              & SI-SDR/dB~$\uparrow$         & WER/\%~$\downarrow$ & PESQ~$\uparrow$               & ESTOI~$\uparrow$              & SI-SDR/dB~$\uparrow$        & WER/\%~$\downarrow$ \\ \midrule
      \multirow{2}{*}{$\lambda=0$}                      & SDE     &          2.58 ± 0.53          &          0.88 ± 0.07          &          14.7 ± 4.2          & 5.10                &          2.68 ± 0.41          &          0.87 ± 0.05          & 6.6 ± 3.7                   &         5.91 \\
                                                        & ODE     &          2.16 ± 0.60          & \textbf{0.90} ± \textbf{0.07} & \textbf{16.3} ± \textbf{4.2} & 5.13                &          2.19 ± 0.49          & \textbf{0.89} ± \textbf{0.04} & \textbf{8.4} ± \textbf{3.5} &         5.29 \\ \midrule
      \multirow{2}{*}{$\ell_1$-norm, $\lambda=10^{-3}$} & SDE     &          2.64 ± 0.55          &          0.89 ± 0.07          &          14.8 ± 4.2          & 4.96                & \textbf{2.77} ± \textbf{0.41} & \textbf{0.89} ± \textbf{0.04} & 7.1 ± 3.9                   & \textbf{4.38} \\
                                                        & ODE     & \textbf{2.81} ± \textbf{0.52} & \textbf{0.90} ± \textbf{0.07} &          16.1 ± 4.1          & \textbf{4.19}       &          2.58 ± 0.46          &          0.85 ± 0.06          & 5.4 ± 4.9                   &         6.29 \\
      \bottomrule
    \end{tabular}
  }
\end{center}
\end{table*}

\subsection{Datasets}
\label{subsec:datasets}
We consider two enhancement tasks: speech denoising and speech dereverberation.
For the speech denoising task, we prepared the \mbox{WSJ0-CHiME3} dataset similarly as in~\cite{lemercier2023storm}.
The dataset was generated using WSJ clean speech~\cite{garofolo_wsj0} and CHiME3 noise~\cite{barker2017third} and the mixture SNR was sampled uniformly in \mbox{[-6, 14]}\,dB~\cite{lemercier2023storm}.
Approximately 13k utterances (25\,h) were generated for the training set, 1.2k utterances (2\,h) for the validation set and 650 utterances (1.5\,h) for the test set.
For the speech dereverberation task, we prepared the \mbox{WSJ0-Reverb} dataset similarly as in~\cite{lemercier2023storm}.
The dataset was generated by convolving WSJ clean speech~\cite{garofolo_wsj0} with room impulse responses (RIRs) simulated using the image method~\cite{scheibler_2018_pyroom}.
Room width and length were sampled uniformly in \mbox{[5, 15]}\,m, and height was sampled in \mbox{[2, 6]}\,m.
Source and microphone locations were selected randomly in the room with a minimum distance of 1\,m from the closest wall.
Reverberation time was sampled in \mbox{[0.4, 1.0]}\,s, and the corresponding anechoic RIR for generating the target signal was generated using a fixed absorption coefficient 0.99.
The sampling rate was 16\,kHz.
\subsection{Experimental setup}
The analysis transform $\mathcal{A}$ is computed using the STFT window size of 510 samples, hop size of 128 samples, and compression parameters $a = 0.5$ and $b = 0.33$~\cite{lemercier2023storm}.
The proposed Schr\"{o}dinger bridge model is denoted as \mbox{SB}, and it is trained using the data prediction loss in~\eqref{eq:data_prediction_loss} with $\lambda=0$, unless stated otherwise, and either VP or VE noise schedule as in Figure~\ref{fig:noise_schedule}.
The VE schedule used $k = 2.6, c = 0.40$, achieving the maximum variance $\sigma_x^2(t)$ of 0.3, twice as large as the maximum OUVE variance of 0.15, similarly as in~\cite{lay_2023_reducing}.
The VP schedule used $\beta_0=0.01, \beta_1=20$ as in~\cite{chen_2023_sb} with variance scale $c=0.3$, achieving the same maximum variance as VE.
Process time for the proposed SB is set to $T = 1$ with $t_\mathrm{min} = 10^{-4}$.
Backbone neural network for the SB models is the noise-conditional score network (\mbox{NCSN++}) proposed in~\cite{song_2021_score} with approximately \mbox{25.2\,M} parameters.
The model is using four down-sampling and up-sampling steps with the configuration following~\cite{lemercier2023storm}.
Training is performed on randomly-selected audio segments corresponding to 256 STFT frames, and the input and the target signals are normalized with the maximum amplitude of the input signal.
The global batch size was set to 64 and the optimizer was Adam with learning rate $10^{-4}$~\cite{lemercier2023storm}.
All models are trained on eight NVIDIA V100 GPUs for a maximum of 1000 epochs with early stopping based on the validation SI-SDR evaluated every 5 epochs and patience of 20 epochs.
We use exponential moving average (EMA) of the weights with decay 0.999~\cite{lemercier2023storm}, and the best EMA checkpoint is selected based on the PESQ value of 50 validation examples~\cite{richter_2023_sgmse,lemercier2023storm}.
Inference is using 50 time steps, unless stated otherwise, with a uniform grid across time.
All models are implemented in NVIDIA's NeMo toolkit~\cite{nvidia_nemo_toolkit}.

The proposed model is compared to three baseline models.
Firstly, we consider a predictive model denoted as \mbox{NSCN++}~\cite{richter_2023_sgmse,lemercier2023storm}, trained to directly estimate the clean speech coefficients $\hat{\mathbf{x}}$ from $\mathbf{y}$~\cite{richter_2023_sgmse, lemercier2023storm}.
Secondly, we consider a diffusion-based generative model denoted as \mbox{SGMSE+}, trained using score matching~\eqref{eq:diffusion_score_matching_loss} with \mbox{NCSN++} as the backbone~\cite{richter_2023_sgmse,lemercier2023storm}.
Thirdly, we consider a hybrid stochastic regeneration model denoted as \mbox{StoRM}~\cite{lemercier2023storm}, consisting of a predictive \mbox{NCSN++} module and a diffusion-based \mbox{SGMSE+} module.
The baseline models were implemented and trained in our framework.
However, the results reported in Section~\ref{subsec:results} were obtained using pre-trained checkpoints from~\cite{lemercier2023storm}, since they were slightly better (approximately 0.05 in PESQ and 0.3\,dB \mbox{SI-SDR} on \mbox{WSJ0-CHiME3}).

The performance is evaluated in terms of perceptual evaluation of speech quality (PESQ)~\cite{rix_2001_pesq}, extended short-term objective intelligibility (ESTOI)~\cite{jensen_2016_estoi}, scale-invariant signal-to-distortion ratio (SI-SDR)~\cite{leroux_2019_sdr} and word error rate (WER).
Clean speech at the microphone was the reference for signal-based metrics.
For WER evaluation we used NVIDIA's \mbox{FastConformer-Transducer-Large} English ASR model~\cite{fastconformer_transducer_large}.
Test examples are available online.\footnote{\url{https://tauaxdefbe.github.io/demo}}
\subsection{Results}
\label{subsec:results}
%
\begin{figure}[bt]
\centering
    \pgfplotstableread[col sep=comma]{tables/num_sampling_steps_unprocessed.csv}\resultsunprocessed
    \pgfplotstableread[col sep=comma]{tables/num_sampling_steps_sgmse.csv}\resultsgmse
    \pgfplotstableread[col sep=comma]{tables/num_sampling_steps_storm.csv}\resultstorm
    \pgfplotstableread[col sep=comma]{tables/num_sampling_steps_sbve_sde.csv}\resultsbvesde
    \pgfplotstableread[col sep=comma]{tables/num_sampling_steps_sbve_ode.csv}\resultsbveode
    \begin{tikzpicture}[baseline,trim axis right]
      \begin{semilogxaxis}[
          width=8.6cm,
          height=3.3cm,
          title=PESQ~$\uparrow$,
          title style={yshift=-1.5ex},
          xmajorgrids=true,
          minor x tick num=9,
          ylabel=PESQ,
          ymajorgrids=true,
          yminorgrids=true,
          minor y tick num=1,
          minor grid style={dotted},
          xtick={3,5,10,20,30,40,50},
          xticklabel=\empty,
          xtick pos=bottom,
          legend style={legend columns=5, nodes={scale=0.7}, at={(0.455,1.55)}, anchor=north, draw=none},
          title style={font=\scriptsize},
          label style={font=\scriptsize},
          tick label style={font=\scriptsize},
          xtick align=inside,
          xmin=3,
          xmax=50,
          ymin=1,
          ymax=3,
          enlarge x limits = {abs=1},
          enlarge y limits = {abs=0.2}
        ]
        \addplot[draw=red, dashed, very thick] table[x=num_steps, y=PESQ]{\resultsunprocessed};
        \addplot[draw=black, very thick, mark=*, mark options={scale=0.8, fill=black}] table[x=num_steps, y=PESQ]{\resultsgmse};
        \addplot[draw=blue, very thick, mark=square*, mark options={scale=0.8, fill=blue}] table[x=num_steps, y=PESQ]{\resultstorm};
        \addplot[draw={rgb:red,118;green,185;blue,0}, very thick, mark=triangle*, mark options={scale=1.3, fill={rgb:red,118;green,185;blue,0}}] table[x=num_steps, y=PESQ]{\resultsbvesde};
        \addplot[draw={rgb:red,118;green,50;blue,0}, dashed, very thick, mark=diamond*, mark options={solid, scale=1.3, fill={rgb:red,118;green,50;blue,0}}] table[x=num_steps, y=PESQ]{\resultsbveode};
        \legend{Unproc., SGMSE+, StoRM, SB-SDE, SB-ODE}
      \end{semilogxaxis}
    \end{tikzpicture}

    \vspace{-0.7em}

    \begin{tikzpicture}[baseline,trim axis right]
      \begin{semilogxaxis}[
          width=8.6cm,
          height=3.3cm,
          title=SI-SDR~$\uparrow$,
          title style={yshift=-1.5ex},
          xmajorgrids=true,
          minor x tick num=9,
          ylabel=SI-SDR / dB,
          ymajorgrids=true,
          yminorgrids=true,
          minor y tick num=1,
          minor grid style={dotted},
          xtick={3,5,10,20,30,40,50},
          xticklabel=\empty,
          xtick pos=bottom,
          title style={font=\scriptsize},
          label style={font=\scriptsize},
          tick label style={font=\scriptsize},
          xtick align=inside,
          xmin=3,
          xmax=50,
          ymin=4,
          ymax=17,
          enlarge x limits = {abs=1},
          enlarge y limits = {abs=1}
        ]
        \addplot[draw=red, dashed, very thick] table[x=num_steps, y=SISDR]{\resultsunprocessed};
        \addplot[draw=black, very thick, mark=*, mark options={scale=0.8, fill=black}] table[x=num_steps, y=SISDR]{\resultsgmse};
        \addplot[draw=blue, very thick, mark=square*, mark options={scale=0.8, fill=blue}] table[x=num_steps, y=SISDR]{\resultstorm};
        \addplot[draw={rgb:red,118;green,185;blue,0}, very thick, mark=triangle*, mark options={scale=1.3, fill={rgb:red,118;green,185;blue,0}}] table[x=num_steps, y=SISDR]{\resultsbvesde};
        \addplot[draw={rgb:red,118;green,50;blue,0}, dashed, very thick, mark=diamond*, mark options={solid, scale=1.3, fill={rgb:red,118;green,50;blue,0}}] table[x=num_steps, y=SISDR]{\resultsbveode};
      \end{semilogxaxis}
    \end{tikzpicture}

    \vspace{-0.7em}

    \begin{tikzpicture}[baseline,trim axis right]
      \begin{semilogxaxis}[
          width=8.6cm,
          height=3.3cm,
          title=WER~$\downarrow$,
          title style={yshift=-1.5ex},
          xlabel=number of steps,
          xlabel style={yshift=1.3ex},
          xmajorgrids=true,
          minor x tick num=9,
          ylabel=WER / \%,
          ymajorgrids=true,
          yminorgrids=true,
          minor y tick num=1,
          minor grid style={dotted},
          xtick={3,5,10,20,30,40,50},
          xticklabels={3,5,10,20,30,40,50},
          xtick pos=bottom,
          title style={font=\scriptsize},
          label style={font=\scriptsize},
          tick label style={font=\scriptsize},
          xtick align=inside,
          xmin=3,
          xmax=50,
          ymin=4,
          ymax=13.5,
          enlarge x limits = {abs=1},
          enlarge y limits = {abs=1}
        ]
        \addplot[draw=red, dashed, very thick] table[x=num_steps, y=WER]{\resultsunprocessed};
        \addplot[draw=black, very thick, mark=*, mark options={scale=0.8, fill=black}] table[x=num_steps, y=WER]{\resultsgmse};
        \addplot[draw=blue, very thick, mark=square*, mark options={scale=0.8, fill=blue}] table[x=num_steps, y=WER]{\resultstorm};
        \addplot[draw={rgb:red,118;green,185;blue,0}, very thick, mark=triangle*, mark options={scale=1.3, fill={rgb:red,118;green,185;blue,0}}] table[x=num_steps, y=WER]{\resultsbvesde};
        \addplot[draw={rgb:red,118;green,50;blue,0}, dashed, very thick, mark=diamond*, mark options={solid, scale=1.3, fill={rgb:red,118;green,50;blue,0}}] table[x=num_steps, y=WER]{\resultsbveode};
      \end{semilogxaxis}
    \end{tikzpicture}
\vspace{-0.7em}
\caption{Speech denoising on \mbox{WSJ0-CHiME3} with different numbers of sampling steps and either \mbox{SB-SDE} or \mbox{SB-ODE} in Table~\ref{table:sb_samplers}. Note that some results for \mbox{SGMSE+} are out of range.}
\label{fig:num_steps}
\end{figure}
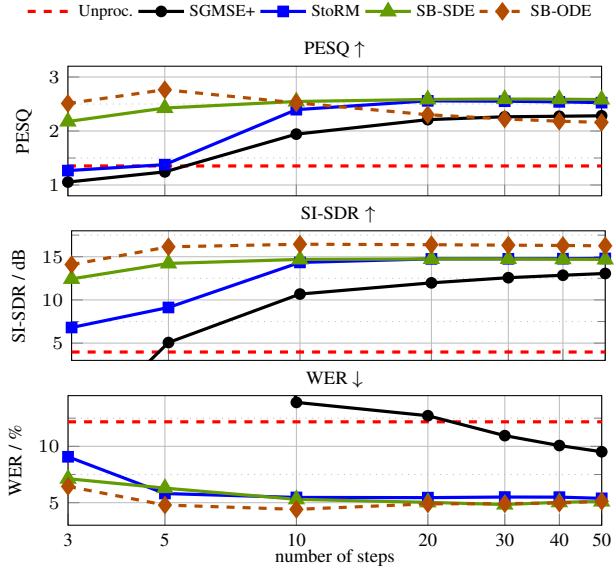

Table~\ref{table:enh} shows the test performance of the models trained on \mbox{WSJ0-CHiME3}.
Predictive \mbox{NCSN++} achieved the best performance in terms of \mbox{SI-SDR}, as also noted in~\cite{richter_2023_sgmse,lemercier2023storm}.
The proposed SB outperformed the diffusion-based \mbox{SGMSE+} model in terms of signal quality and ASR performance by a large margin.
Furthermore, SB performed better or on par with the hybrid \mbox{StoRM} model, without using a separate predictive model.
In general, SB resulted in significantly less hallucinations that \mbox{SGMSE+} and performed similarly to \mbox{StoRM}.
\mbox{SB-VE} and \mbox{SB-VP} performed similarly in terms of signal quality metrics, with the latter achieving a better ASR performance.

Table~\ref{table:der} shows the test performance of the models trained on \mbox{WSJ0-Reverb}.
The proposed \mbox{SB-VE} outperformed the diffusion-based \mbox{SGMSE+} model in terms of signal quality, while achieving a similar ASR performance.
The proposed \mbox{SB-VE} performed better than the hybrid \mbox{StoRM} model in signal quality metrics, although it lagged in ASR evaluation.
In this task, \mbox{SB-VP} performed worse than \mbox{SB-VE} in all metrics.
Since \mbox{SB-VE} performed well in both tasks, it was used for the rest of the experiments.

Table~\ref{table:aux_loss} shows the performance of the \mbox{SB-VE} model with different samplers and with the auxiliary loss.
With data prediction loss ($\lambda=0$), the ODE sampler performed well in terms of \mbox{ESTOI}, \mbox{SI-SDR} and WER.
However, the SDE sampler performed significantly better in terms of PESQ.
The tradeoff parameter $\lambda$ in~\eqref{eq:data_prediction_loss} was selected from $10^{\left\lbrace -4, \dots, 1 \right\rbrace}$ and $\lambda = 10^{-3}$ showed a good validation performance.
With $\mathcal{L}_\text{aux}$, the results for both SDE and ODE sampler were mostly improved across the board.
The best result on \mbox{WSJ0-CHiME3} is achieved using the ODE sampler, outperforming the best baseline \mbox{StoRM} in Table~\ref{table:enh} in all metrics, resulting in a relative WER reduction of more than 20\%.
The best result on \mbox{WSJ0-Reverb} is achieved using the SDE sampler, outperforming the best baseline \mbox{StoRM} in Table~\ref{table:enh} in all metrics, resulting in a relative WER reduction of more than 6\%.

Finally, the influence of the number of steps for the reverse process for \mbox{WSJ0-CHiME3} is investigated in Figure~\ref{fig:num_steps}.
SB is using the same \mbox{SB-VE} model and only the sampler is changed at inference time.
In can be observed that SB is more robust to the number of steps used in the reverse process, performing significantly better than the baseline diffusion \mbox{SGMSE+} and the hybrid \mbox{StoRM} models.
The performance gap is especially large for WER, where \mbox{SGMSE+} performs poorly as the number of steps is reduced.
Furthermore, SB performs better than the hybrid \mbox{StoRM} model, especially for a small number of steps, without a separate predictive model.
Interestingly, \mbox{SB-ODE} shows an improved performance in PESQ and WER for a small number of steps, while still performing well in \mbox{SI-SDR}.
Note that the baseline \mbox{SGMSE+} and StoRM models are using the same number of steps, but they employ a predictor-corrector sampler, which performs two calls to backbone neural networks per step~\cite{richter_2023_sgmse,lemercier2023storm}.
The proposed SB models with samplers from Table~\ref{table:sb_samplers} perform only one call to the backbone neural network per step, resulting in a significant reduction in computational complexity.
\section{Conclusions}
In this paper, we presented a speech enhancement model based on the Schr\"{o}dinger bridge.
As opposed to diffusion-based models, the proposed model is based on a data-to-data process.
The proposed model outperforms both diffusion-only baseline and a hybrid baseline, combining predictive and generative models, both in terms of signal quality and ASR performance.
In general, very good performance has been observed across the tested conditions, e.g., with relative WER reduction of more than 20\% for denoising and 6\% for dereverberation compared to the best baseline model.
Furthermore, the proposed SB model is more robust to the number of steps used in the reverse process, performing significantly better than the baseline models.
\FloatBarrier
%
\cleardoublepage
\bibliographystyle{template/IEEEtran}
\bibliography{main}

\end{document}